\documentclass[12pt]{article}       
\usepackage{graphicx}

\begin{document}

\title{Noncommutativity in the early Universe}
              
\author{G. Oliveira-Neto, M. Silva de Oliveira\footnote{The first two authors are from: Departamento de F\'{\i}sica,
Instituto de Ci\^{e}ncias Exatas, 
Universidade Federal de Juiz de Fora,
CEP 36036-330 - Juiz de Fora, MG, Brazil. gilneto@fisica.ufjf.br and monalisa-silva@hotmail.com}, G. A. Monerat\footnote{Departamento de Modelagem Computacional,
Instituto Polit\'{e}cnico do Rio de Janeiro,
Universidade do Estado do Rio de Janeiro,
Rua Bonfim, 25 - Vila Am\'{e}lia - Cep 28.625-570,
Nova Friburgo, RJ, Brazil.
monerat@uerj.br}\\ and
        E. V. Corr\^{e}a Silva\footnote{Faculdade de Tecnologia,
Universidade do Estado do Rio de Janeiro,
Rodovia Presidente Dutra, Km 298, P\'{o}lo Industrial,
CEP 27537-000, Resende-RJ, Brazil. evasquez@uerj.br}}

\maketitle

\begin{abstract}
In the present work, we study the noncommutative version of a quantum
cosmology model. The model has a Friedmann-Robertson-Walker geometry, the 
matter content is a radiative perfect fluid and the spatial sections have zero constant curvature. 
In this model the scale factor takes values in a bounded domain. Therefore, 
its quantum mechanical version has a discrete energy spectrum. We compute the discrete 
energy spectrum and the corresponding eigenfunctions. The energies depend
on a noncommutative parameter $\beta$. We compute the scale factor expected 
value ($\left<a\right>$) for several values of $\beta$. For all of them, $\left<a\right>$
oscillates between maxima and minima values and never vanishes. It gives an initial
indication that those models are free from singularities, at the quantum level. We improve
this result by showing that if we subtract a quantity proportional to the standard deviation of $a$ 
from $\left<a\right>$, this quantity is still positive. 
The $\left<a\right>$ behavior, for the present model, is a drastic 
modification of the $\left<a\right>$ behavior in the corresponding commutative version of the present model. 
There, $\left<a\right>$ grows without limits with the time variable. 
Therefore, if the present model may represent the early stages of the Universe, the results of the present 
paper give an indication that $\left<a\right>$ may have been, initially, bounded due to noncommutativity.
We also compute the Bohmian trajectories for $a$,
which are in accordance with $\left<a\right>$, and the quantum potential $Q$. From $Q$, we may
understand why that model is free from singularities, at the quantum level.
\end{abstract}

\maketitle

\section{Introduction}
\label{introduction}

The idea of noncommutative degrees of freedom was first introduced, in physics, 
a long time ago, by Snyder \cite{snyder,snyder1}. 
There, the noncommutativity was imposed between the spacetime coordinates 
and his main motivation was to eliminate the divergences in quantum field theory.
Recently, the interest in those noncommutativity ideas were 
renewed due to some important results obtained in superstring, membrane and $M$-theories 
\cite{banks,connes,chu,schomerus,witten}. For more information on those
important results we refer to the reviews \cite{douglas,szabo}. Since then, 
noncommutativity has been applied to many other physical systems, such as: quantum harmonic oscillator
\cite{nair,gamboa,bellucci}, hydrogen atom \cite{chaichian}, quantum Hall effect
\cite{dayi,dayi1,kokado}, Einstein's gravity theory \cite{chamseddine,meyer,meyer1}, cosmology 
\cite{brandenberger,huang,kim,liu,huang1,kim1,nelson0,nozari,pedram,obregon,neves,gil},
black hole physics \cite{nicolini,nicolini1,rizzo,nicolini2,zet,nicolini3,banerjee1,nicolini4,banerjee2,brown},
quantum cosmology \cite{garcia,nelson,barbosa,gil1}, to name only but a few. For a more complete list of references 
see \cite{banerjee}.

One important arena where noncommutative (NC) ideas may play an important role 
is cosmology. In the early stages of its evolution, the Universe may have had
very different properties than the ones it has today. Among those properties
some physicists believe that the spacetime coordinates were subjected to a 
noncommutative algebra. Inspired by these ideas some researchers have considered 
such NC models in quantum cosmology 
\cite{garcia,nelson,barbosa,gil1}. It is also possible that some residual NC contribution 
may have survived in later stages of our Universe. Based on these ideas some 
researchers have proposed some NC models in classical cosmology in order to explain 
some intriguing results observed by WMAP. Such as a running spectral index of the 
scalar fluctuations and an anomalously low quadrupole of CMB angular power spectrum 
\cite{huang,kim,liu,huang1,kim1}. Another relevant application of the NC ideas in 
semi-classical and classical cosmology is the attempt to explain the present accelerated 
expansion of our Universe \cite{pedram,obregon,neves,gil}.

In Ref. \cite{gil}, several different noncommutative classical 
Friedmann-Robertson-Walker (FRW) cosmological models were studied. There, they
work in Schutz's variational formalism \cite{schutz,germano1} and use the Hamiltonian 
formalism. Therefore, the phase space of those models is given by the following canonical 
variables and conjugated momenta: $\{ a, p_a, T, p_T \}$, where $a$ is the scale factor,
$T$ is a time variable associated to the fluid and $p_a$ and $p_{T}$ are, respectively,
their conjugated momenta. They consider a noncommutativity relation between the two momenta $p_a$ and
$p_{T}$. In subsection 4.3, page 15 of Ref. \cite{gil}, they consider a model, that may represent
the early stages of our Universe, with flat spatial 
sections and a radiative perfect fluid. For a positive noncommutative parameter, they show that 
the scale factor behavior is drastically modified with respect to the corresponding commutative 
version of the model. For the commutative version, the scale factor grows and eventually goes to 
infinity when the time goes to infinity, following the equation (in the gauge $N=1$) \cite{dinverno},
\begin{equation}
\label{0}
a(t) = \sqrt{\sqrt{4E/3}\, t + a_0^2},
\end{equation}
where $t$ is the time coordinate, $E$ is the radiation energy and $a_0$ is the scale factor value for $t=0$.
On the other hand, in the noncommutative version the scale factor remains bounded. If the Universe starts expanding
from a small scale factor value, after a finite time it reaches a maximum value and then contracts to the singularity. 
In order to see that behavior, consider equations (4.2), (4.3) and (4.11) of Ref. \cite{gil}. From them, we obtain the 
following equations describing the scale factor dynamics (in the gauge $N=1$),
\begin{equation}
\label{0.5}
\dot{a}^2 + \frac{\beta}{3a} - \frac{E}{3a^2} = 0,
\end{equation}
\begin{equation}
\label{0.6}
2\ddot{a}a + \dot{a}^2 + \frac{E}{3a^2} = 0,
\end{equation}
where the dot means derivative with respect to the coordinate time $t$, $\beta$ is the noncommutative parameter and we have 
used the notation of the present paper to name the fluid energy ($E$). If one chooses $\beta=0.1$,
$E=1.336713605$, $a(t=0)=0.1$ and $\dot{a}(t=0)=6.650096751$ and solve Eqs. (\ref{0.5}-\ref{0.6}), one obtains the result 
shown in Figure 1. In that figure, the scale factor stops before reaching the singularity due to numerical limitations. 
It means that, if this model may represent the early stages of the Universe, it gives an indication that the scale 
factor may have been, initially, bounded due to noncommutativity. Since, quantum cosmology is more appropriate to explain the initial stages of 
the Universe, than classical cosmology, we have decided investigating if that important indication is still true, at the
quantum level.

\begin{figure}
\centerline{\includegraphics[width=7cm,height=5cm, angle=0]{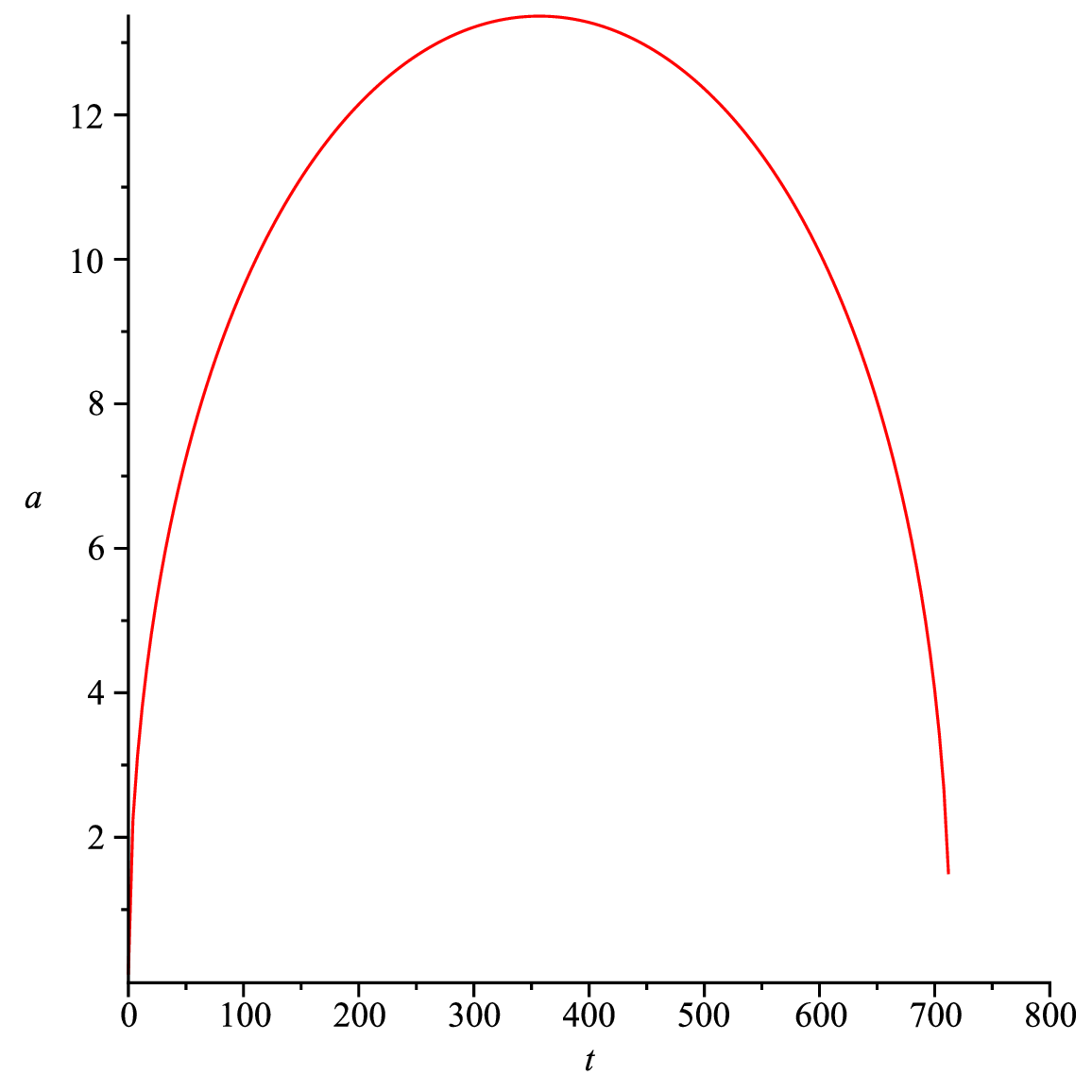}}
\vspace*{8pt}
\caption{$a(t)$ as a function of $t$, for $\beta=0.1$, $E=1.336713605$, $a(t=0)=0.1$ and $\dot{a}(t=0)=6.650096751$.\protect\label{fig1}}
\end{figure}

In the present work, we study the quantum cosmology version of the noncommutative model described above. 
The noncommutativity, at the quantum level, we are about to propose will be between the canonically conjugated 
momenta to the scale factor and the radiative perfect fluid, following the choice made, at the classical level, 
by the authors of Ref. \cite{gil}. Since these variables are functions of the time coordinate $t$, this procedure is 
a generalization of the typical noncommutativity between usual spatial coordinates. The noncommutativity between 
those types of phase space variables have already been proposed in the literature. At the quantum
level in Refs. \cite{garcia,nelson,barbosa,gil1} and at the semi-classical and classical levels
in Refs. \cite{pedram,obregon,neves,gil}. We quantize the model and obtain the appropriate 
Wheeler-DeWitt equation. In this model the scale factor takes values in a bounded domain. Therefore, its 
quantum mechanical version has a discrete energy spectrum. We compute the discrete 
energy spectrum and the corresponding eigenfunctions. The energies grow with
a noncommutative parameter $\beta$. We compute the scale factor expected 
value ($\left<a\right>$) for several values of $\beta$. For all of them, $\left<a\right>$
oscillates between maxima and minima values and never vanishes. It gives an initial
indication that those models are free from singularities, at the quantum level. We improve
this result by showing that if we subtract a quantity proportional to the standard deviation of $a$ 
from $\left<a\right>$, this quantity is still positive. We observe that, 
$\left<a\right>$ grows with the decrease of $\beta$. We also observe that,
the smaller the value of $\beta$, the greater is the interval where $\left<a\right>$ takes values. All
these results confirm, at the quantum level, the results obtained in Ref. \cite{gil}, for the scale factor, 
at the classical level. We also compute the Bohmian trajectories for $a$,
which are in accordance with $\left<a\right>$, and the quantum potential $Q$. From $Q$, we may
understand why that model is free from singularities, at the quantum level.

In the next section, we obtain the Wheeler-DeWitt equation for the NC model and solve it. The wavefunction is a
linear combination of products of Airy functions and time exponentials. The number of terms contributing to the
wavefunction is given by $N$. We compute the $\left<a\right>$ as a function of the NC parameter $\beta$ and $N$.
We also compute, $\left<a\right>-\alpha\sigma_a$, where $\sigma_a$ stands for the standard deviation of $a$ and 
$\alpha$ is a real number. We show that for certain values of $\alpha$ this quantity is always positive, which
improves the result that $\left<a\right>$ never goes to zero. In Section \ref{debroglie-bohm}, we compute the 
Bohmian trajectories for $a$ and show that they are in accordance with $\left<a\right>$. We also compute the 
quantum potential $Q$. Studying $Q$, we show why that model is free from singularities, at the quantum level.
Finally, in Section \ref{conclusions}, we discuss the most important results of the present paper.

\section{Quantum Cosmology in the Many Worlds Intepretation}
\label{many worlds}

The FRW cosmological models are characterized by the
scale factor $a(t)$ and have the following line element,
\begin{equation}  
\label{1}
ds^2 = - N^2(t) dt^2 + a^2(t)\left( \frac{dr^2}{1 - kr^2} + r^2 d\Omega^2
\right)\, ,
\end{equation}
where $d\Omega^2$ is the line element of the two-dimensional sphere with
unitary radius, $N(t)$ is the lapse function and $k$ gives the type of
constant curvature of the spatial sections. Here, we are considering the 
case with zero curvature $k=0$ and we are using the natural
unit system, where $\hbar=c=G=1$. The matter content of the model is
represented by a perfect fluid with four-velocity $U^\mu = \delta^{\mu}_0$
in the comoving coordinate system used. The total energy-momentum tensor 
is given by,
\begin{equation}
T_{\mu,\, \nu} = (\rho+p)U_{\mu}U_{\nu} - p g_{\mu,\, \nu}\, ,  
\label{2}
\end{equation}
where $\rho$ and $p$ are the energy density and pressure of the fluid,
respectively. Here, we assume that $p = \rho/3$, which is the equation of
state for radiation. This choice may be considered as a first approximation
to treat the matter content of the early Universe and it was made as a
matter of simplicity. It is clear that a more complete treatment should
describe the radiation, present in the primordial Universe, in terms of the
electromagnetic field.

From the metric (\ref{1}) and the energy momentum tensor (\ref{2}), one may 
write the total Hamiltonian of the present model ($N {\mathcal{H}}$), where
$N$ is the lapse function and ${\mathcal{H}}$ is the superhamiltonian constraint.
It is given by \cite{germano1},

\begin{equation}
N {\mathcal{H}}= -\frac{p_{a}^2}{12} + p_{T},  
\label{3}
\end{equation}
where $p_{a}$ and $p_{T}$ are the momenta canonically conjugated to $a$ and 
$T$, the latter being the canonical variable associated to the fluid \cite
{schutz,germano1}. Here, we are working in the conformal gauge, where $N = a$. The 
commutative version of the present model was first treated in Ref. \cite{lemos}.

We wish to quantize the model following the Dirac formalism for quantizing
constrained systems \cite{dirac}. First we introduce a wave-function which
is a function of the canonical variables $a$ and $T$,

\begin{equation}  
\label{4}
\Psi\, =\, \Psi(a ,T )\, .
\end{equation}
Then, we impose the appropriate commutators between the operators $a$
and $T$ and their conjugate momenta $p_a$ and $p_T$.
Working in the Schr\"{o}dinger picture, the operators $a$ and $T$
are simply multiplication operators, while their conjugate momenta are
represented by the differential operators,
\begin{equation}
p_{a}\rightarrow -i\frac{\partial}{\partial a}\hspace{0.2cm},\hspace{0.2cm} 
\hspace{0.2cm}p_{T}\rightarrow -i\frac{\partial}{\partial T}\hspace{0.2cm}.
\label{5}
\end{equation}

Finally, we demand that the operator corresponding to $N \mathcal{H}$ 
annihilate the wave-function $\Psi$, which leads to the Wheeler-DeWitt 
equation,
\begin{equation}
\frac{1}{12}\frac{{\partial}^2}{\partial a^2}
\Psi(a,\tau) = -i \, \frac{\partial}{\partial \tau}\Psi(a,\tau),
\label{6}
\end{equation}
where the new variable $\tau= -T$ has been introduced. This is the Schr\"{o}dinger 
equation of an one dimensional free particle restricted to the positive domain of 
the variable.

The operator $N \hat{\mathcal{H}}$ is self-adjoint \cite{lemos} with respect
to the internal product,

\begin{equation}
(\Psi ,\Phi ) = \int_0^{\infty} da\, \,\Psi(a,\tau)^*\, \Phi (a,\tau)\, ,
\label{7}
\end{equation}
if the wave functions are restricted to the set of those satisfying either 
$\Psi (0,\tau )=0$ or $\Psi^{\prime}(0, \tau)=0$, where the prime $\prime$
means the partial derivative with respect to $a$. Here, we consider wave 
functions satisfying the former type of boundary condition and we also 
demand that they vanish when $a$ goes to $\infty$. For the boundary conditions 
mentioned above, the author of Ref. \cite{lemos} solved Eq. (\ref{6}) and used 
that solution to compute $\left<a\right>$ for that model. He obtained, for the
boundary condition $\Psi (0,\tau )=0$ (in the gauge $N=a$),
\begin{equation}
\label{7.5}
\left<a\right> = \frac{1}{6} \sqrt{\frac{2}{\pi\sigma}} \sqrt{\sigma^2 \tau^2 + (6 - p \tau)^2},
\end{equation}
where $\sigma$ is a positive number, $p$ is a real number and $\tau$ is the time variable. 
Therefore, $\left<a\right>$ starts from a nonzero value and when $\tau$ grows 
it also grows. Eventually, when $\tau\to \infty$ also $\left<a\right> \to \infty$. From Eq. (\ref{7.5}), 
in that limit, $\left<a\right> \propto \tau$.

In order to introduce the noncommutativity in the present model, we shall
modify the prescription used in Refs. \cite{garcia,nelson,barbosa,gil1}. In those models the 
noncommutativity was described by a non-zero commutator between the operators associated to the 
canonical variables $a$ and $T$. Here, the non-zero commutator will be between the two operators 
associated to the canonical momenta $p_a$ and $p_T$,
\begin{equation}
\label{8}
\left[ \breve{p_a}, \breve{p_T} \right] = i\beta\, ,
\end{equation}
where $\breve{p_a}$ and $\breve{p_T}$ are the noncommutative versions of the
operators and $\beta$ is the positive NC parameter. We follow, here, the choice compatible, at the quantum level, with the one made by the authors of 
Ref. \cite{gil}, at the classical level. This noncommutativity between those operators can be taken
to functions that depend on the noncommutative version of those operators with the 
aid of the Moyal product \cite{moyal,bayen,witten,douglas}. Consider 
two functions of $\breve{a}$ and $\breve{T}$, 
let's say, $f$ and $g$. Then, the Moyal product between those two function is given by: 
$f(\breve{a},\breve{T}) \star g(\breve{a},\breve{T}) = f(\breve{a},\breve{T})
\exp{\left[(i\theta/2)(\overleftarrow{\partial_{\breve{a}}}\overrightarrow{\partial_{\breve{T}}} -
\overleftarrow{\partial_{\breve{T}}}\overrightarrow{\partial_{\breve{a}}})\right]}g(\breve{a},\breve{T})$.

Using the Moyal product, we may adopt the following Wheeler-DeWitt equation
for the noncommutative version of the present model,
\begin{equation}
\label{9}
\left[\frac{1}{12}\breve{p}_{a}\star\breve{p}_{a} - \breve{p}_{T}\right] \star \Psi(\breve{a},\breve{T}) = 0.
\end{equation}

It is possible to rewrite the Wheeler-DeWitt equation (\ref{9}) in terms of a commutative version
of the operators $\breve{p_a}$ and $\breve{p_T}$ and the ordinary product of functions. In order to do that,
we must initially introduce the following transformation between the noncommutative and the commutative 
operators,

\begin{eqnarray}
\label{10}
\breve{p_a} & = & p_a + \beta T,\\
\breve{p_T} & = & p_T,\nonumber
\end{eqnarray}
and the transformations of the other noncommutative variables are trivial: $\breve{a}=a$ and $\breve{T}=T$.
We follow, here, the choice compatible, at the quantum level, with the one made by the authors of Ref. \cite{gil},
at the classical level. Then, we may write the commutative version of the Wheeler-DeWitt equation (\ref{9}), to first 
order in the commutative parameter $\beta$, in the Schr\"{o}dinger picture as,

\begin{equation}
\label{12}
\frac{1}{12} \frac{\partial^2 \Psi(a,\tau)}{\partial a^2} 
-\frac{i}{6} \beta \tau \frac{\partial \Psi(a,\tau)}{\partial a}
= - i \frac{\partial \Psi(a,\tau)}{\partial \tau},
\end{equation}
where we have made the following transformation $\tau \to -T$, in the same way the author of Ref. \cite{lemos},
so that, we may compare the $\left<a\right>$ computed using our solution with Eq. (\ref{7.5}).
For a vanishing $\beta$ this equation reduces to the commutative Schr\"{o}dinger equation (\ref{6}), described above.

In order to solve this equation, satisfying the boundary conditions: $\Psi (0,\tau )=0$ and $\lim_{a\to\infty} \Psi (a,\tau )\to 0$, 
we start imposing that the wave function $\Psi(a,\tau)$ has the following form,
\begin{equation}
\label{13}
\Psi(a,\tau) = e^{i\beta a \tau} e^{-iE\tau} A(a).
\end{equation}
Introducing this ansatz in  Eq. (\ref{12}), we obtain, to first order in $\beta$, the eigenvalue equation,

\begin{equation}
\label{14}
\frac{d^2 A(a)}{da^2} - (12\beta a - 12E)A(a) = 0,
\end{equation}
where $E$ is the eigenvalue and it is associated with the fluid energy. 

The solutions to this equation are the Airy functions,
\begin{displaymath}
\label{15}
A(a) = c_{1} Ai\left(
\frac{12\theta a - 12E}{(12\beta)^{2/3}}\right)+
c_{2} Bi\left(
\frac{12\theta a - 12E}{(12\beta)^{2/3}}\right).
\end{displaymath}
The Airy functions $Bi$ grow up exponentially when $a\rightarrow \infty$. 
In order to eliminate this undesirable behavior, we put $c_{2}=0$. 
Then, the energy eigenfunctions for our model are,
\begin{equation}
\label{16}
A(a)= c_{1} Ai\left(
\frac{12\beta a - 12E}{(12\beta)^{2/3}}\right).
\end{equation}
If we introduce the boundary condition that $A(a=0)=0$, we find from Eq. (\ref{16}) the
energy eigenvalues with the following expression,
\begin{equation}
\label{17}
E_n = \frac{1}{12} (12\beta)^{2/3} \alpha_n
\end{equation}
where $\alpha_n$ is positive and is the zero of order $n$ of the Airy function $Ai$. It is clear from
this equation that the energy eigenvalues grow with $\beta$.

The most general expression of $\Psi(a,\tau)$ Eq. (\ref{13}), which is a solution to Eq. (\ref{12}), is a 
linear combination of the eigenfunctions $A_n(a)$, Eq. (\ref{16}), taking in account the energy eigenvalues
Eq. (\ref{17}), combined with the exponential factor present in Eq. (\ref{13}), for a given $\beta$ value.

\begin{equation}
\Psi(a,\tau)= e^{i\beta a \tau/2}\sum_{n=0}^{N} C_n Ai\left(
\frac{12\beta a - 12E_{n}}{(12\beta)^{2/3}}\right)\exp{(-iE_{n}\tau)}.
\label{18}
\end{equation}

In order to build a wave packet from Eq. (\ref{18}) one has, initially, to fix the values of $\beta$
and the number $N$ of energy eigenfunctions contributing to the sum. After that, one has to compute the $N$ 
energy eigenvalues $E_n$ Eq. (\ref{17}), with the aid of the first $N$ zeros ($\alpha_n$) of the Airy function $Ai$.
Also, one has to fix the values of the $N$ coefficients $C_n$. Finally, one has to introduce the 
explicit values of all those quantities in Eq. (\ref{18}) and perform the indicated sum.
The time evolution of the wave packets built from Eq. (\ref{18})
shows that they are null not only at the origin but they are asymptotically 
null at infinity as well. In the region near $a=0$ these packets present strong 
oscillations, which decrease as $a$ increases. 

Now, we shall use the wavefunction (\ref{18}) in order to compute some important quantities. 
Initially, we shall compute, the scale factor expected value, $\left<a\right>$, for different
values of $\beta$ and $N$. First of all, let us choose $C_n=1$, for all $n$, in Eq. (\ref{18}).
In fact, we shall do this choice for the $C_n$'s coefficients in all calculations in this paper.
Next, we compute the eigenvalues, $E_n$, with the aid of Eq. (\ref{17}). In order to do that we must 
choose the values of $\beta$, $N$ and $\alpha_n$. In the present situation, we shall choose several different 
values of $\beta$ and $N$. Finally, we must compute the scale factor expected value, using the following expression,

\begin{equation}
\label{19}
\left<a\right> = \frac{\int_{0}^{\infty}a\,|\Psi (a,\tau)|^2 da} {
\int_{0}^{\infty}|\Psi (a,\tau)|^2 da}.
\end{equation}

After computing $\left<a\right>$ for several different values of $\beta$, $N$ and various $\tau$ intervals, we 
noticed that this quantity oscillates between maxima and minima values and never vanishes. It gives an initial
indication that those models are free from singularities, at the quantum level. Now, if we fix $N$ and vary $\beta$
we observe the following properties of $\left<a\right>$: ($i$) the maximum value of $\left<a\right>$ decreases with the increase of $\beta$; ($ii$) the
amplitude of oscillation for $\left<a\right>$ decreases with the increase of $\beta$; ($iii$) the number of $\left<a\right>$ 
oscillations, for a fixed $\tau$ interval, increases with the increase of $\beta$.  
Those behaviors may be understood by the fact that the potential barrier, that confines 
the scale factor, grows linearly with $\beta$. Therefore, as $\beta$ increases the $\left<a\right>$ is forced to oscillate in an ever 
decreasing region. Under those conditions, for fixed $N$, the maximum value and the amplitude of $\left<a\right>$ decrease. Also,
since the domain where $\left<a\right>$ oscillates is decreasing, the number of $\left<a\right>$ 
oscillations, for a fixed $\tau$ interval, increases. All those properties can be seen in Figures 2 and 3. Each figure shows the behavior 
of $\left<a\right>$ for a different value of $\beta$ while the $\tau$ interval and $N$ remain fixed. Now, if we fix $\beta$ and vary $N$
we observe the following properties of $\left<a\right>$: ($i$) the maximum value of $\left<a\right>$ grows with the increase of $N$; ($ii$) the
amplitude of oscillation for $\left<a\right>$ increases with the increase of $N$; ($iii$) the number of $\left<a\right>$ 
oscillations, for a fixed $\tau$ interval, decreases with the increase of $N$. In order to understand those behaviors we notice that the mean
energy associated with the wavepacket increases with the increase of $N$. Therefore, for fixed $\beta$, when we increase $N$ the domain where
$\left<a\right>$ oscillates increases. In this way, the maximum value and the amplitude of $\left<a\right>$ increase. On the other hand,
the number of $\left<a\right>$ oscillations, for a fixed $\tau$ interval, decreases.
All those properties can be seen in Figures 4 and 5.

As we have mentioned above, for all values of $\beta$ and $N$ considered, $\left<a\right>$ never vanishes. It gives an initial
indication that those models are free from singularities, at the quantum level. We may improve this result by computing $\left<a\right>-\alpha\sigma_a$,
where $\sigma_a$ stands for the standard deviation of $a$ and $\alpha$ is a positive real number. If this quantity is always positive like $\left<a\right>$, it will be a
stronger indication that the model is free from singularities, at the quantum level. Let us compute $\left<a\right>-\alpha\sigma_a$, for the present model.
By definition the standard deviation of $a$ is given by,

\begin{equation}
\label{20}
\sigma_a = \sqrt{\left<a^2\right> - \left<a\right>^2},
\end{equation}
where,
\begin{equation}
\label{21}
\left<a^2\right> = \frac{\int_{0}^{\infty}a^2\,|\Psi (a,\tau)|^2 da} {
\int_{0}^{\infty}|\Psi (a,\tau)|^2 da},
\end{equation}
and $\left<a\right>^2$ is given by the square of Eq. (\ref{19}). Using the wavefunction (\ref{19}) and repeating some procedures we did in order to
compute $\left<a\right>$, we computed $\left<a\right>-\alpha\sigma_a$ for several values of $\alpha$, $\beta$ and $N$. The result is that for the huge majority
of cases this quantity is always positive. More precisely, if $\alpha \leq 3/4$, in the interval $10^{-7} \leq \beta \leq 0.5$, $\left<a\right>-\alpha\sigma_a$ is always
positive for any value of $N$. As for the mathematical significance of $\alpha=3/4$, we may mention that if our distribution were a normal one and if one takes the interval 
$\left<a\right> \pm \alpha\sigma_a$, around the mean value, it would cover over half the area under the distribution. More precisely, $54,67\%$ \cite{meyer2}.  Two examples of 
$\left<a\right>-3\sigma_a/4$, as a function of time, are shown in Figures 6 and 7.

Therefore, we notice that the introduction of the noncommutativity represented by
Eq. (\ref{8}) modified in an important way the commutative version of the model. In the commutative version of the model
the scale factor expected value takes values in an unbounded domain. It expands as the function of $\tau$ given by Eq. (\ref{7.5}). 
On the other hand, in the noncommutative version of the model the scale factor mean value takes values 
in a bounded domain and is periodic in $\tau$. The commutative version of the model may be obtained from the noncommutative 
one by taking the limit when $\beta \to 0$. The above results show clearly that limit from one version to the other.
If we start decreasing the value of $\beta$ the scale factor expected value will oscillate in an ever increasing domain until we
set $\beta \to 0$. At that limit Eq. (\ref{12}) reduces to Eq. (\ref{6}) and the scale factor expected value will grow without 
limits, as the function of $\tau$ given by Eq. (\ref{7.5}).

\begin{figure}[ph]
\centerline{\includegraphics[width=7cm,height=5cm, angle=0]{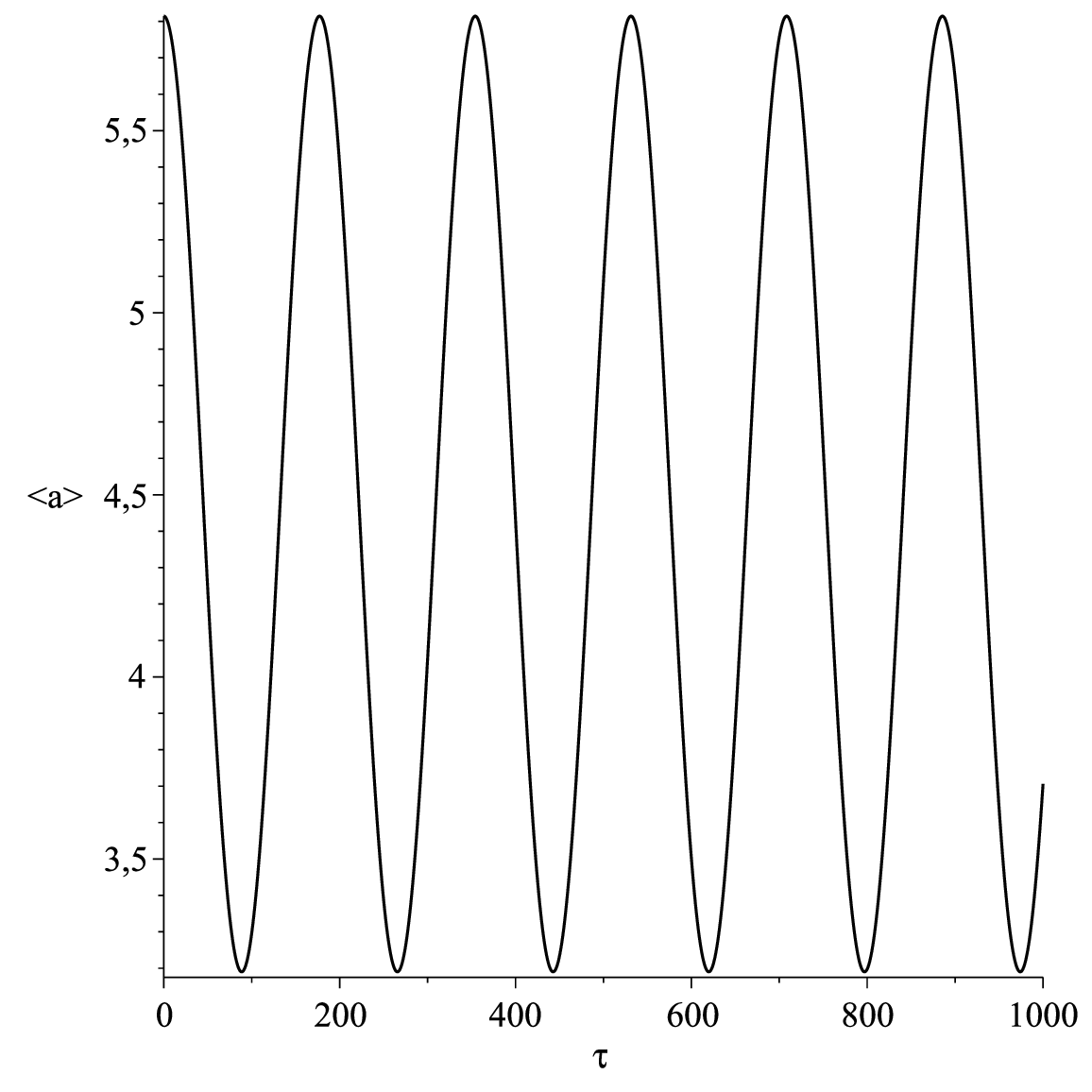}}
\vspace*{8pt}
\caption{$\left<a\right>$ for $\beta=0.01$, $N=2$ and the time interval $0 \leq \tau \leq 1000$.\protect\label{fig2}}
\end{figure}

\begin{figure}[ph]
\centerline{\includegraphics[width=7cm,height=5cm, angle=0]{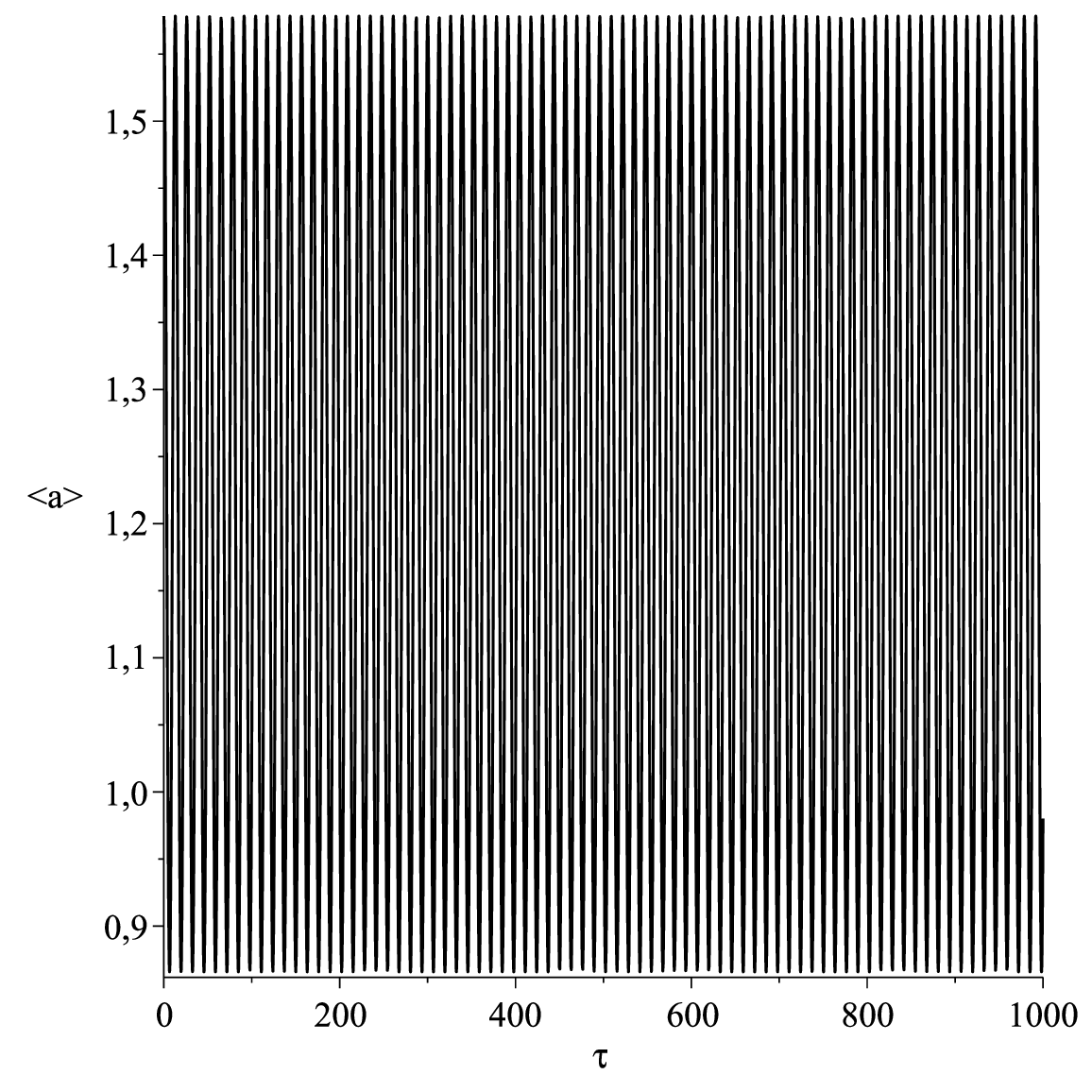}}
\vspace*{8pt}
\caption{$\left<a\right>$ for $\beta=0.5$, $N=2$ and the time interval $0 \leq \tau \leq 1000$.\protect\label{fig4}}
\end{figure}

\begin{figure}[ph]
\centerline{\includegraphics[width=7cm,height=5cm, angle=0]{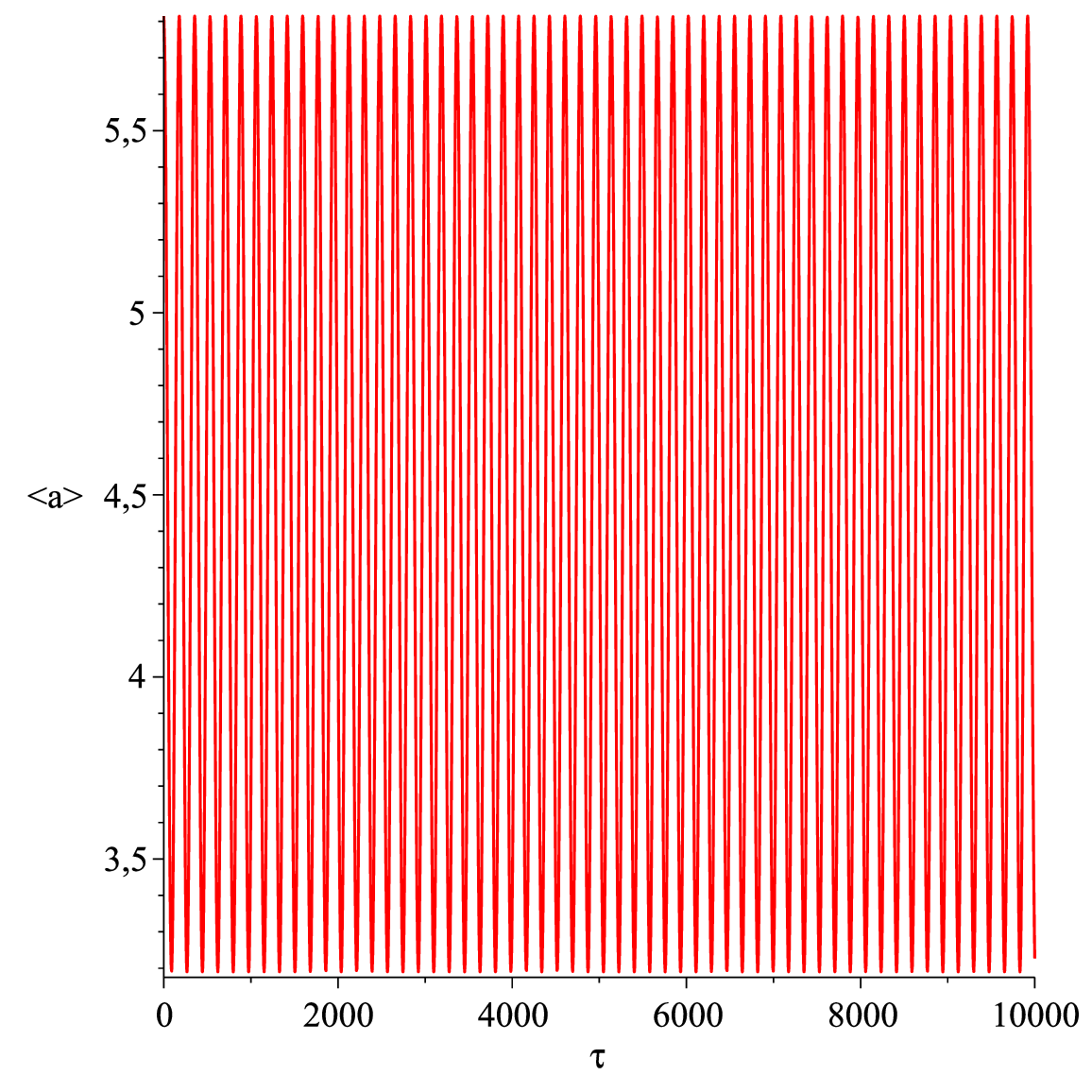}}
\vspace*{8pt}
\caption{$\left<a\right>$ for $N=2$, $\beta=0.01$ and the time interval $0 \leq \tau \leq 10000$.\protect\label{fig5}}
\end{figure}

\begin{figure}[ph]
\centerline{\includegraphics[width=7cm,height=5cm, angle=0]{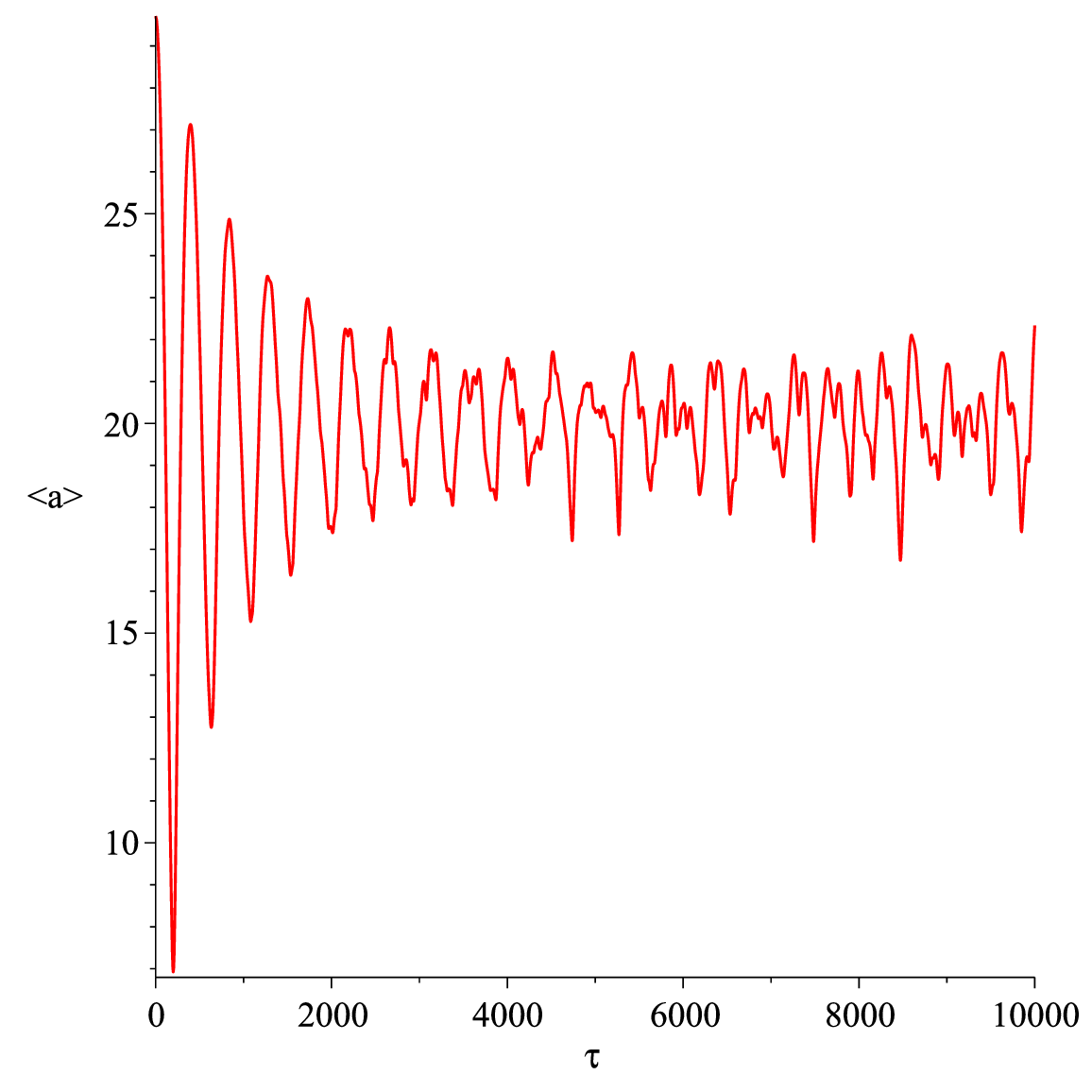}}
\vspace*{8pt}
\caption{$\left<a\right>$ for $N=22$, $\beta=0.01$ and the time interval $0 \leq \tau \leq 10000$.\protect\label{fig6}}
\end{figure}

\begin{figure}[ph]
\centerline{\includegraphics[width=7cm,height=5cm, angle=0]{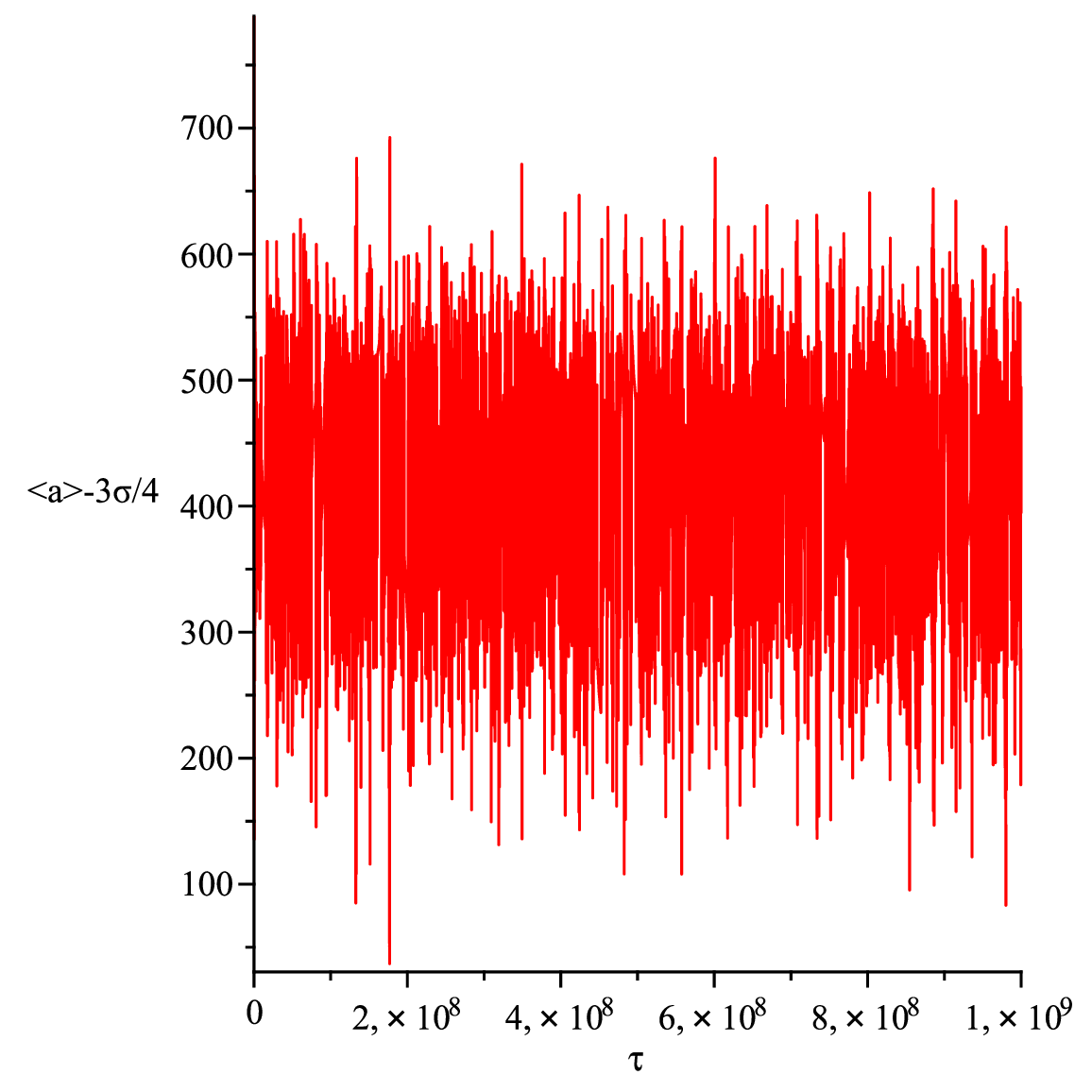}}
\vspace*{8pt}
\caption{$\left<a\right>-3\sigma_a/4$ for $N=15$, $\beta=0.0000001$ and the time interval $0 \leq \tau \leq 10^9$.\protect\label{fig8}}
\end{figure}

\begin{figure}[ph]
\centerline{\includegraphics[width=7cm,height=5cm, angle=0]{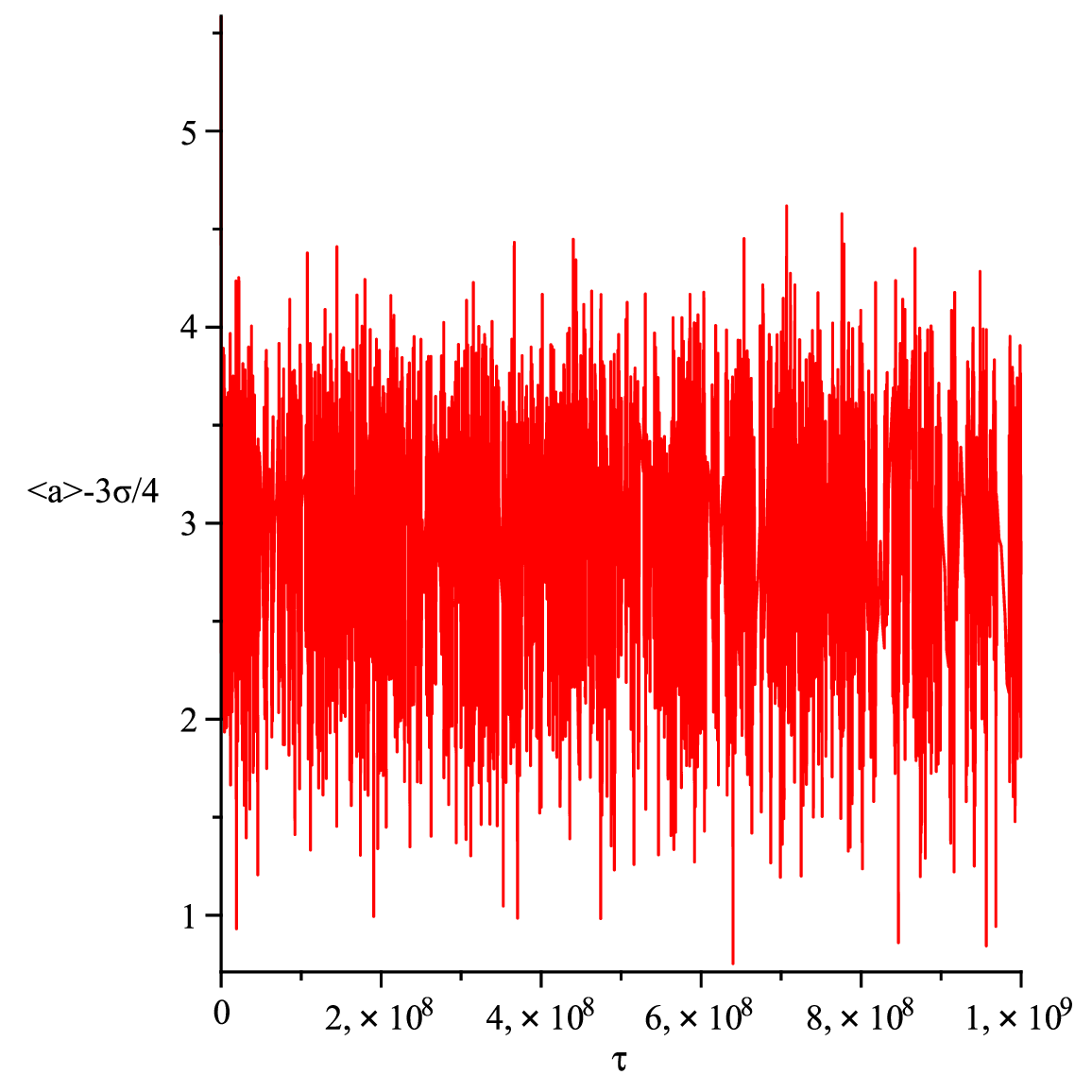}}
\vspace*{8pt}
\caption{$\left<a\right>-3\sigma_a/4$ for $N=20$, $\beta=0.5$ and the time interval $0 \leq \tau \leq 10^9$.\protect\label{fig10}}
\end{figure}

\section{Quantum Cosmology in the DeBroglie-Bohm Intepretation}
\label{debroglie-bohm}

In this section, we want to apply the {\it DeBroglie-Bohm} interpretation
of quantum mechanics, to the present NC quantum cosmology model. Our main
motivation is to compare the results we shall obtain with that interpretation 
with the ones we obatined in the previous section, where
we used the {\it Many Worlds} interpretation of quantum mechanics.
In order to use the {\it DeBroglie-Bohm} interpretation we must
re-write $\Psi (a,\tau)$ Eq. (\ref{18}), in the polar form,
\begin{equation}
\Psi(a,\tau)=R(a,\tau) e^{iS(a,\tau)}
\label{22}
\end{equation}
where,
\begin{equation}
R(a,\tau)=
\sqrt{\sum_{n,m=0}^{N} C_n C_m Ai\left( G \left( \beta,E_n,a \right)\right)
Ai\left( G \left( \beta,E_m,a \right)\right)\cos\left((E_{n}-E_{m})\tau\right)}
\label{23}
\end{equation}

\begin{equation}
S(a,\tau)=\arctan\left[
\frac{-\sum_{n=0}^{N} C_n Ai\left( G \left( \beta,E_n,a \right)\right)\sin(E_{n}\tau)}
{\sum_{m=0}^{N} C_m Ai\left( G \left( \beta,E_m,a \right)\right)\cos(E_{m}\tau)}
\right].
\label{24}
\end{equation}

\noindent
where $G \left( \beta,E_n,a \right)=
\left(12\beta a - 12E_{n}\right)/(12\beta)^{2/3}$.

Following the {\it DeBroglie-Bohm} interpretation we introduce 
$\Psi (a,\tau)$ Eq. (\ref{22}) in Eq. (\ref{12}), this leads to the next two
equations for $R (a,\tau)$ and $S(a,\tau)$ \cite{holland},

\begin{eqnarray}
\label{25}
12\frac{\partial S(a,\tau)}{\partial \tau} - 2\beta\tau\frac{\partial S(a,\tau)}{\partial a} +
\left(\frac{\partial S(a,\tau)}{\partial a}\right)^2 +
Q(a,\tau)&=&0,\\ \nonumber
& & \\
\frac{\partial R(a,\tau)}{\partial \tau} +
\frac{1}{6}\frac{\partial S(a,\tau)}{\partial a}\frac{\partial R(a,\tau)}{\partial a} +
\frac{1}{12}R(a,\tau)\frac{\partial^2 S(a,\tau)}{\partial a^2} - \frac{1}{6}\beta\tau\frac{\partial R(a,\tau)}{\partial a}&=&0,\\ \nonumber
\end{eqnarray}
where the Bohmian quantum potential $Q(a,\tau)$ is defined by \cite{holland},
\begin{equation}
\label{26}
Q(a,\tau)=-\frac{1}{R(a,\tau)}\frac{\partial^2 R(a,\tau)}{\partial a^2}.
\end{equation}

In the present situation, using the value of $R (a,\tau)$ 
Eq. (\ref{23}), $Q (a,\tau)$ Eq. (\ref{26}) takes the form,

\begin{equation}
\label{27}
Q(a,\tau) = \frac{1}{4} \frac{1}{M_1^2} \left( \frac{\partial M_1}{\partial a}\right)^2 - \frac{1}{2}\frac{1}{M_1}\frac{\partial^2 M_1}{\partial a^2},
	\end{equation}
	
\noindent
where

\begin{equation}
M_1 = \sum_{n,m=0}^{N} C_n C_m \ Ai\left(G(\beta, E_n, a)\rule{0mm}{4mm}\right) Ai\left(G(\beta, E_m, a)\rule{0mm}{4mm}\right) \cos\left((E_n - E_m)\tau\rule{0mm}{4mm}\right).\end{equation}

The Bohmian trajectory for $a$ is given by \cite{holland},
\begin{equation}
\label{28}
\frac{da(\tau)}{d\tau}=\frac{1}{m}\frac{\partial S}{\partial a},
\end{equation}
where, from Eq. (\ref{3}) $m$ for the present situation is given by $6$.
Using the value of $S(a,\tau)$ Eq. (\ref{24}) in Eq. (\ref{28}), it
reduces to,

\begin{equation}
\label{29}
	\frac{da(\tau)}{d\tau} = \frac{1}{6}
	\frac{F_3(\tau)}{F_4(\tau)}
	\end{equation}	
where,
{\small
\begin{eqnarray}
	F_3(\tau) &=& \sum_{n, m=0}^{N}
		C_{n} C_{m}  
		{Ai}^{\prime}\left( G \left( \beta,E_{{n}},a \left( \tau \right)\right)\rule{0mm}{4mm}\hspace{-1mm}\right)
		{Ai}\left( G \left( \beta,E_{{m}},a \left( \tau \right) \right)\rule{0mm}{4mm}\hspace{-1mm}\right) \times \nonumber \\
		& & \times 
		 \sin \left( \left( E_{{n}}-E_{{m}} \right)\tau\rule{0mm}{4mm}\right),
	\end{eqnarray}

\begin{eqnarray}
	F_4(\tau) &=& \left[ \sum _{n=0}^{N} C_{n}
		 {Ai}\left( G \left( \beta,E_{{n}},a \left( \tau \right)\right)\rule{0mm}{4mm}\hspace{-1mm}\right) 
		\cos \left( E_{n}\tau \right)  \right] ^{2} + \nonumber \\
		& & 
		 \left[\sum_{m=0}^{N} C_{m}
		 Ai \left( G \left( \beta,
		E_{{m}},a \left( \tau \right)\right)\rule{0mm}{4mm}\hspace{-1mm}\right) 
		\sin \left( E_{{m}}\tau \right)  \right] ^{2}.
	\end{eqnarray}	}

The solution to Eq. (\ref{29}), which is the Bohmian 
trajectory of $a(\tau)$, which is the variable describing the
universe, represents the quantum behavior for the cosmic 
evolution in the Planck era.

We solved Eq. (\ref{29}) for many different values of 
$\beta$ and $N$, the number of energy eigenfunctions contributing to the 
wavefunction Eq. (\ref{22}). We found the same qualitative behavior 
for the Bohmian trajectories of $a(\tau)$, in all those cases. 
It oscillates between maxima and minima values and never goes
through the zero value. It means that, quantum mechanically, in
those models there are no singularities which confirms the
result obtained in the previous section using the {\it Many Worlds} interpretation.
In order to exemplify this behavior we show the Bohmian trajectories
of $a(\tau)$ for two models with $N=2$ and $\beta = 0.01$, Figure 8, and $\beta = 0.5$, Figure 9. 
We computed the time evolution of $a(\tau)$ up to 
$t=1000$ and used the initial conditions for $a(\tau)$ at $\tau=0$,
obtained from the calculation of the expected 
value of $a(\tau)$, for the corresponding models. The results shown in Figures 
8 and 9 are qualitatively very similar to Figures 2 and 3, that represent
the scale factor expected values for the corresponding models. For different values
of $N$, the behavior of $a(\tau)$ is also qualitatively very similar to the behavior
of $\left<a\right>$. As an example, we show $a(\tau)$, Figure 10, for the model where $N=22$,
$\beta=0.01$, $0\leq \tau \leq 10000$ and the initial condition for $a(\tau)$ at $\tau=0$ was 
obtained from the calculation of $\left<a\right>$. This Figure must be compared with Figure 5, 
for $\left<a\right>$ of the corresponding model.

\begin{figure}[ph]
\centerline{\includegraphics[width=7cm,height=5cm, angle=0]{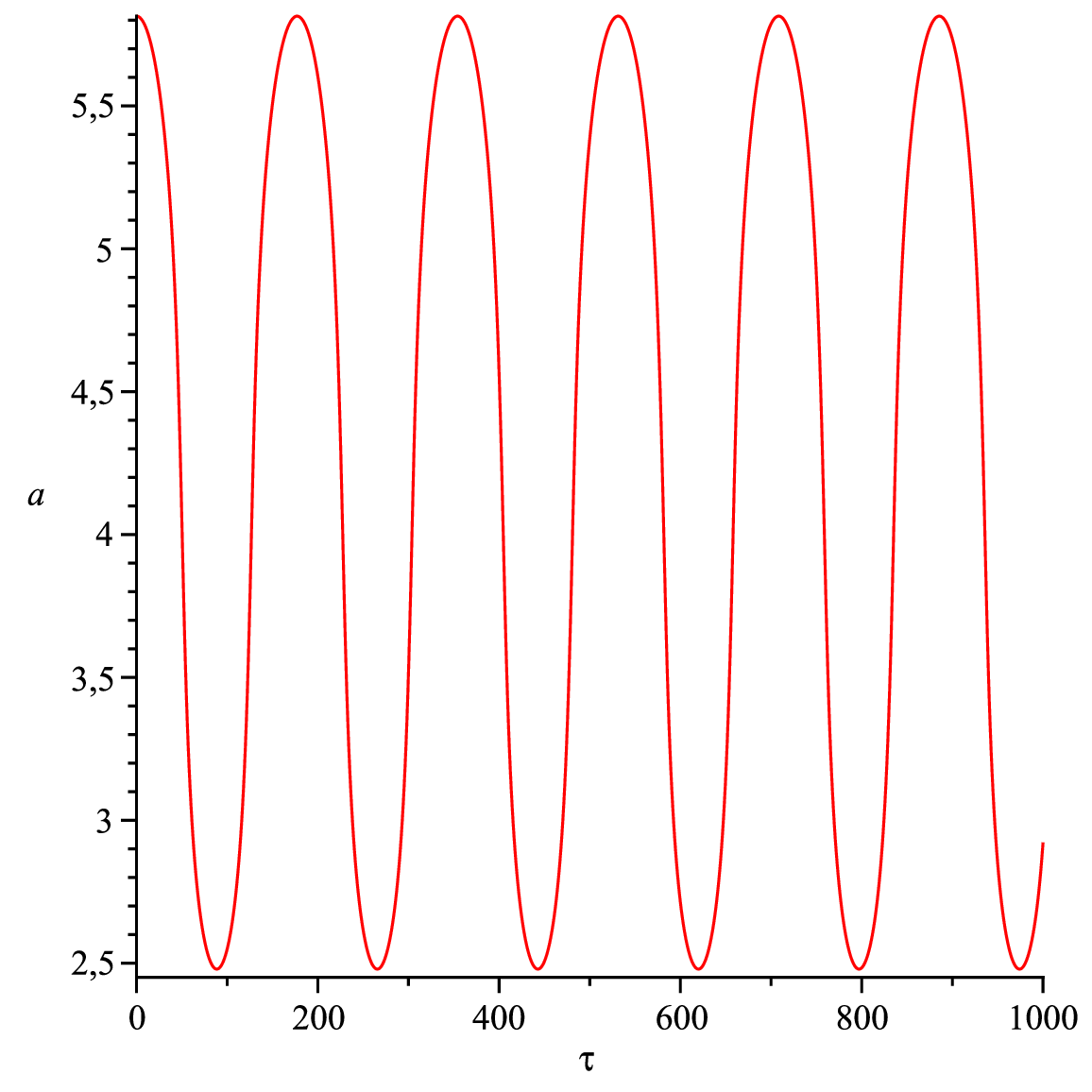}}
\vspace*{8pt}
\caption{$a(\tau)$ for $N=2$, $\beta=0.01$, the initial condition $a(\tau=0)=5.814364999$ and the time interval $0 \leq \tau \leq 1000$.\protect\label{fig11}}
\end{figure}

\begin{figure}[ph]
\centerline{\includegraphics[width=7cm,height=5cm, angle=0]{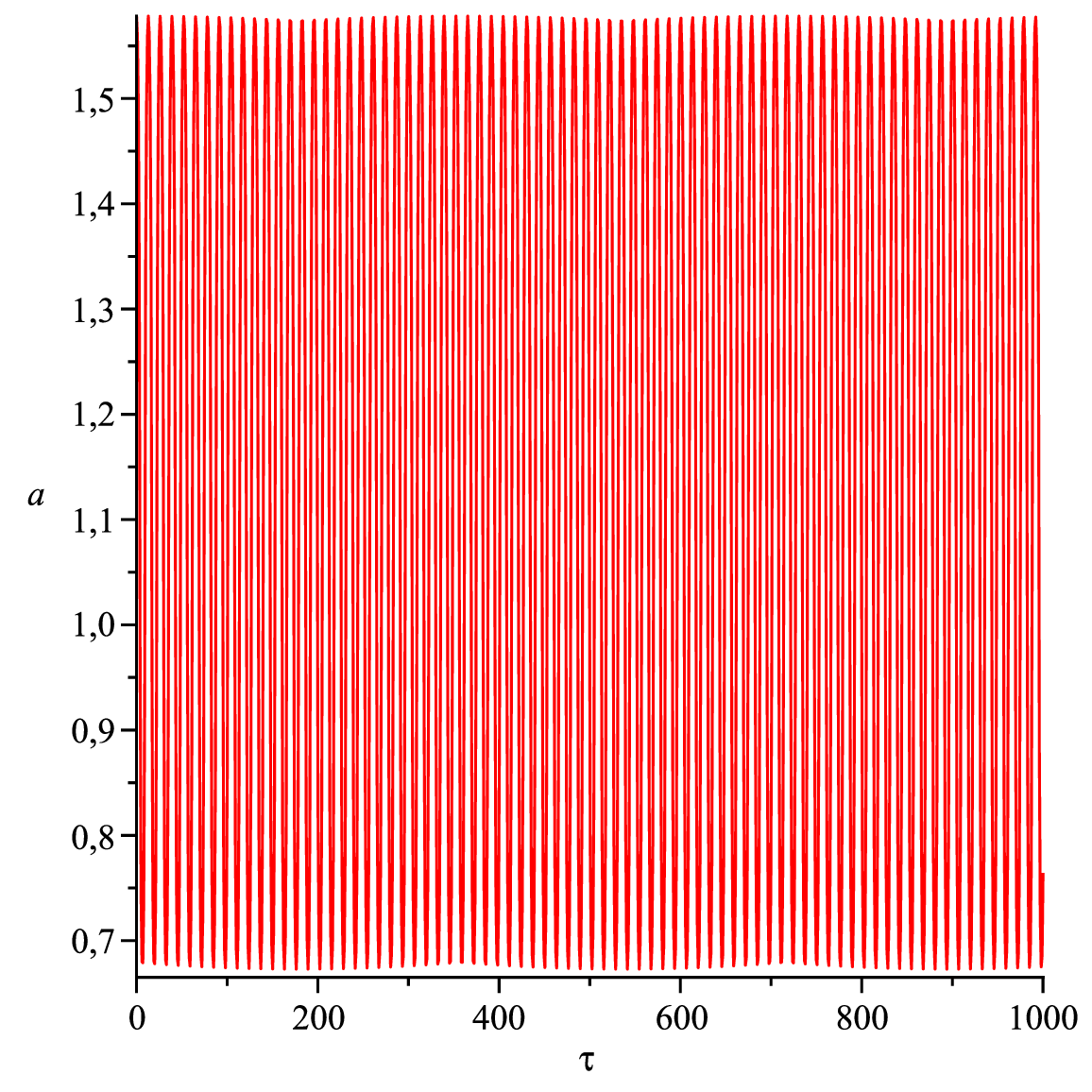}}
\vspace*{8pt}
\caption{$a(\tau)$ for $N=2$, $\beta=0.5$, the initial condition $a(\tau=0)=1.578261478$ and the time interval $0 \leq \tau \leq 1000$.\protect\label{fig13}}
\end{figure}

\begin{figure}[ph]
\centerline{\includegraphics[width=7cm,height=5cm, angle=0]{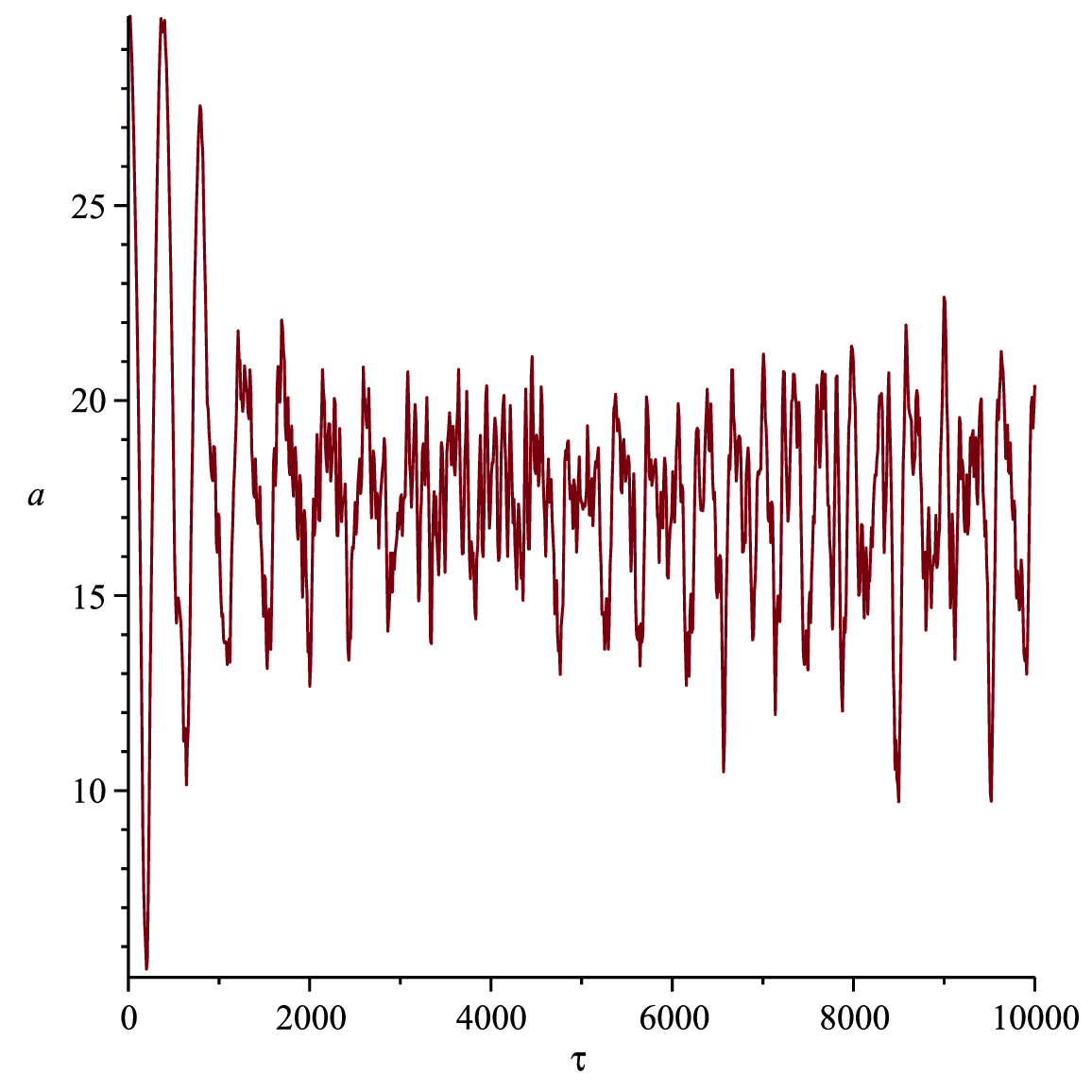}}
\vspace*{8pt}
\caption{$a(\tau)$ for $N=22$, $\beta=0.01$, the initial condition $a(\tau=0)=29.71117629$ and the time interval $0 \leq \tau \leq 10000$.\protect\label{fig14}}
\end{figure}

The absence of singularities in the present models are very easy
to understand when one observes the Bohmian quantum potential Eq. (\ref{27}),
for those models. We computed $Q(a,\tau)$ Eq. (\ref{27}), for several 
values of $\beta$ and $N$. 
The calculations were made over the Bohmian trajectories of $a$. We obtained
$Q$ as a function of $\tau$ as well as a function of $a$. We found the same 
qualitative behavior of $Q$, in all those cases. Initially, considering $Q$ as a function of $\tau$, at $\tau=0$, there is
a potential barrier ($B_0$) that prevents the value of $a$ ever to go through zero. 
Then, the barrier becomes a well for a brief moment and again a new barrier 
appears ($B_1$). After a while, $B_1$ turns into a well for a brief moment
and then another barrier identical to $B_0$ appears. After that, $Q$,
periodically, repeats itself. $B_0$ is different from $B_1$. $B_1$ exists for a 
longer period and is shorter than $B_0$. One may interpret the potential shape in 
the following way. Initially, at $\tau=0$, $a$ starts to grow from its minimum value 
different from zero, first rapidly, and then its velocity starts to decrease until 
it goes to zero, at the maximum value of $a$. Then, $a$ starts to decrease, first 
slowly, and then its velocity starts to increase until $a$ reaches its minimum value 
different from zero. There, its velocity 
changes sign and $a$ starts to grow once more, as described above. This dynamics is 
represented in $Q$, initially, by $B_0$, then the first well, then $B_1$ and finally
the well just after $B_1$. Then, the movement of $a$ repeats itself periodically.
These models have no singularities because $B_0$ and its periodic repetitions
prevent $a$ ever to go through zero. In order to exemplify this behavior we show, in
Figure 11, the Bohmian quantum potential Eq. (\ref{27}), for the model with 
$\beta = 0.1$ and $N=2$. For a better visualization of $Q$'s
behavior, we choose a small time interval in Figure 11. We computed, also, $Q$ as a function of $a$.
In this case, we may see clearly $B_0$ and $B_1$. In Figure 12, we show $Q$ as a function of $a$, for
the model with $\beta = 0.1$ and $N=2$. For a clearer understand
of $Q$'s behavior we plotted, in Figure 13, the Bohmian trajectory of $a$
used in order to compute $Q$ given in Figures 11 and 12. $a$, Figure 13, is plotted during the same
time interval of $Q$, Figure 11, and its initial condition $a(\tau)$ at $\tau=0$, was obtained
from the calculation of the expected value of $a$, for the same model.

\begin{figure}[ph]
\centerline{\includegraphics[width=7cm,height=5cm, angle=0]{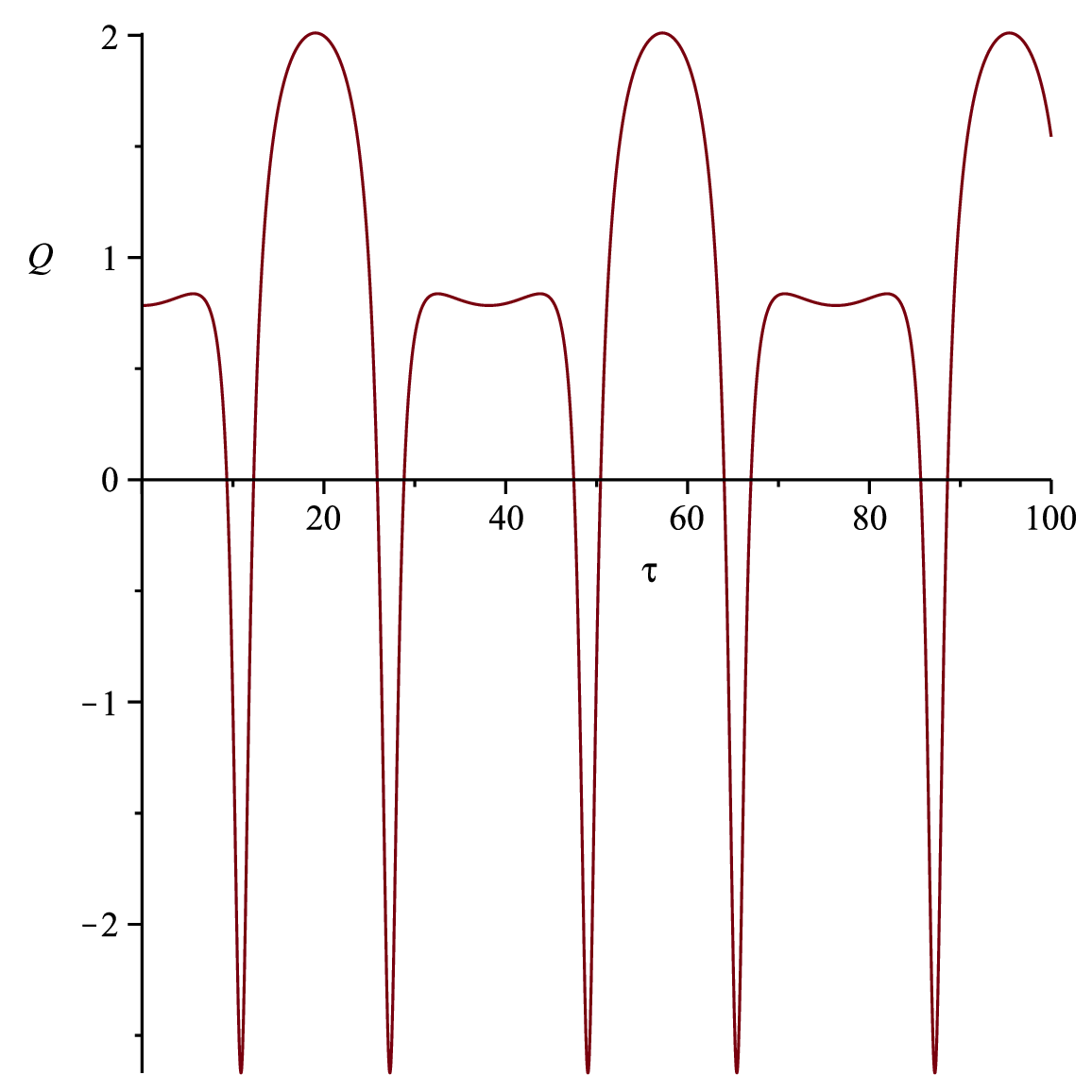}}
\vspace*{8pt}
\caption{$Q(\tau)$ for $N=2$, $\beta=0.1$ and the time interval $0 \leq \tau \leq 100$.\protect\label{fig15}}
\end{figure}

\begin{figure}[ph]
\centerline{\includegraphics[width=7cm,height=5cm, angle=0]{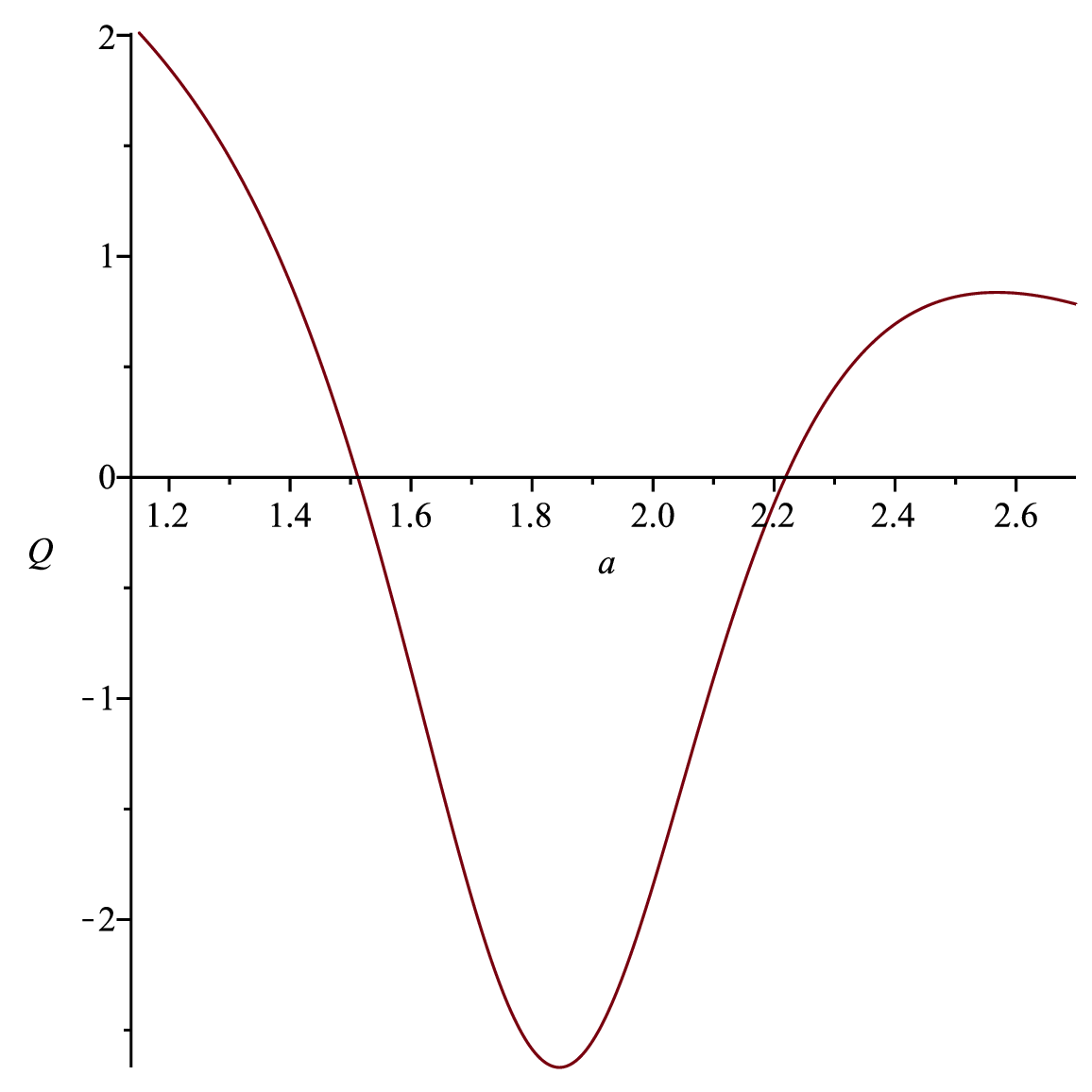}}
\vspace*{8pt}
\caption{$Q(a)$ for $N=2$, $\beta=0.1$ in the $a$ interval {$1.150667559 \leq a \leq 2.698789166$}.\protect\label{fig16}}
\end{figure}

\begin{figure}[ph]
\centerline{\includegraphics[width=7cm,height=5cm, angle=0]{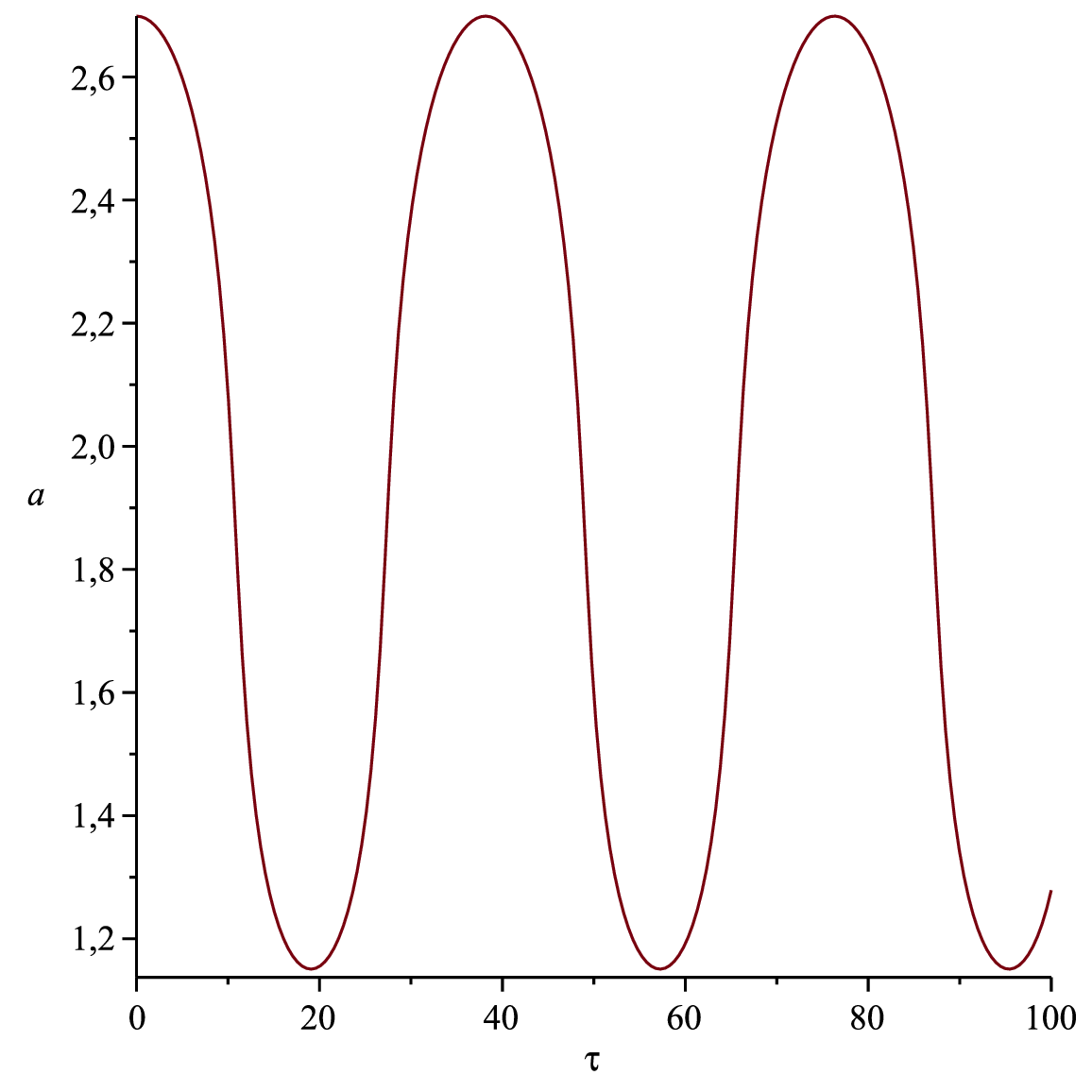}}
\vspace*{8pt}
\caption{$a(\tau)$ for $N=2$, $\beta=0.1$, the initial condition $a(\tau=0)=2.698789166$ and the time interval $0 \leq \tau \leq 100$.\protect\label{fig17}}
\end{figure}

\section{Conclusions}
\label{conclusions}

The above results indicate that, also at the quantum level, the scale factor may have been,
initially, bounded, due to the presence of noncommutativity. The difference between the scale 
factor behavior at the classical and the quantum levels is the fact that, in the quantum NC 
version of the model both the scale factor expected value and its Bohmian trajectory oscillate 
between maxima and minima values and never go to zero. Therefore, this NC cosmological model is 
free from singularities, at the quantum level. In the many words interpretation, it is possible 
to improve this result by showing that the quantity $\left<a\right>-\alpha\sigma_a$ is always 
positive for many values of $\alpha$. Where $\sigma_a$ stands for the standard deviation of $a$ 
and $\alpha$ is a positive real number. From the quantum potential $Q$, for that NC model, it is 
easy to see why the Bohmian trajectory for $a$ never goes to zero. On the other hand, in the 
classical NC version of the present model, the scale factor starts expanding from a minimum value, 
then reachs a maximum value and finally contracts to zero, giving rise to a singularity. It is 
important to investigate if other types of noncommutativity 
applied to models with the conditions present in our early Universe, also give rise to scale factors 
which take values in a bounded domain. That would indicate an important prediction of NC cosmological
models about the early universe. In this sense, we may mention that in a previous work \cite{gil1} 
the authors quantized a noncommutative FRW model with $k=1$ and radiation. For a noncommutativity described 
by a non-zero commutator between the scale factor ($a$) and the variable associated to the radiative fluid 
($T$), they showed that it is not possible to solve the Wheeler-DeWitt equation for that model and find a 
wavefunction, with the boundary condition $\Psi (0,T)=0$.

{\bf Acknowledgements}. M. Silva de Oliveira thanks CAPES for her scholarship.
G. A. Monerat thank UERJ for the Prociencia grant.

\end{document}